%% file: fit.tex
\documentclass[copyright,creativecommons]{eptcs}
\providecommand{\event}{Foundations of Interface Technologies 2012} 
\usepackage{url}
\usepackage{breakurl}             
\usepackage{amsmath,amssymb,amsthm}
\usepackage[T1]{fontenc}
\usepackage{graphicx,subfigure}
\usepackage{epstopdf}
\usepackage[all]{xy}
\usepackage[inference,ligature]{semantic}
\usepackage{todonotes}

\definecolor{citeblue}{rgb}{0.4,0,.6}
\definecolor{refcolor}{rgb}{0,0.1,0.5}
\hypersetup{pdftex,colorlinks=true,bookmarks=false,linkcolor=refcolor,citecolor=citeblue,urlcolor=blue}

\include{macros}
\title{Refinement for Transition Systems with Responses\thanks{
Supported by Danish Council for Strategic Research grants 2106-080046 The
Jingling Genies, 2106-07-0019
Trustworthy Pervasive Healthcare Services, and by Villum
Kann Rasmussen Foundation through VKR Centre of Excellence MT-LAB. Authors listed alphabetically.} }
\author{Marco Carbone \quad Thomas Hildebrandt 
\quad Gian
Perrone \quad Andrzej W\k{a}sowski
\institute{IT University of Copenhagen, Denmark}
\email{\{maca,hilde,gdpe,wasowski\}@itu.dk}
}
\newcommand{\titlerunning}{Refinement for Transition Systems with Responses}
\newcommand{\authorrunning}{Carbone. Hildebrandt.  Perrone.
W\k{a}sowski}

\begin{document}
\maketitle

\begin{abstract}
Motivated by the response pattern for property specifications and applications within flexible workflow management systems, 
we report upon an initial study of modal and mixed transition systems in which the must transitions are interpreted as must \emph{eventually}, and in which implementations can contain may behaviors that are resolved at \emph{run-time}.
We propose Transition Systems with Responses (TSRs) as a suitable model for this study. We prove that TSRs correspond to a restricted class of mixed transition systems, which we refer to as the 
action-deterministic mixed transition systems. 
   We show that TSRs 
allow for a natural definition of deadlocked and accepting states. We then transfer the standard definition of refinement for mixed transition systems to TSRs and prove that refinement does not preserve deadlock freedom. 
This leads to the proposal of \emph{safe}  refinements, which are 
those that preserve deadlock freedom. We exemplify the use of TSRs and (safe) refinements on a small medication workflow. 
\end{abstract}

\input newintro

\section{Action-deterministic Modal and Mixed Transition Systems}
\label{sec:modalmixed}\input modalmixed

\section{Transition Systems with Responses and Refinement}
\label{sec:tsr}\input tsr

\section{Conclusion and Future Work}
\label{sec:future}
\input conclusion

\bibliographystyle{eptcs}
\bibliography{shortbib}
\end{document}

%% file: macros.tex
\mathlig{|->}{\mapsto}
\mathlig{-->}{\longrightarrow}
\mathlig{->}{\rightarrow}
\mathlig{=>}{\Rightarrow}
\mathlig{==>}{\Longrightarrow}
\mathlig{||-}{\Vdash}
\mathlig{|-}{\vdash}
\mathlig{|=}{\models}
\mathlig{||}{\parallel}
\mathlig{[|}{[\![}
\mathlig{|]}{]\!]}
\mathlig{[]}{\Box}
\mathlig{<<>>}{\diamondsuit}
\mathlig{<=io}{\leq_{io}}
\mathlig{|>}{\rhd}
\mathlig{<|}{\lhd}
\mathlig{<.}{\langle}
\mathlig{.>}{\rangle}
\def\actionSet{\mathsf{Act}}
\newcommand{\action}[1]{\xrightarrow{#1}} 
\newcommand{\may}{<<>>}
\newcommand{\must}{\square}
\newcommand{\power}{{\cal{P}}}
\newcommand{\deadlock}{\mathit{deadlock}}
\newtheorem{lemma}{Lemma}
\newtheorem{definition}{Definition}
\newtheorem{proposition}{Proposition}
\newtheorem{theorem}{Theorem}

\newcommand{\cat}[1]{\ensuremath{\mathbf{#1}}}
\newcommand{\DMixTS}{\ensuremath{\cat{DMixTS}_0}}
\newcommand{\DMTS}{\ensuremath{\cat{DMTS}_0}}
\newcommand{\TSR}{\ensuremath{\cat{TSR}_0}}
\newcommand{\MTSR}{\ensuremath{\cat{MTSR}_0}}
\newcommand{\obj}[1]{\ensuremath{\cat{#1}_0}}

%% file: newintro.tex
\section{Introduction}\label{sec:intro}

Modal transition systems (MTS) were introduced originally in the seminal work
of Larsen and Thomsen\,\cite{DBLP:conf/lics/LarsenT88} (see also
\cite{antonik.ea:2008:beatcs}) as a basic transition system model supporting
stepwise specification and refinement of parallel processes.  Intuitively, an MTS can be 
considered a labeled transition system (LTS) in which a subset of the
transitions are identified as being required (must), while the others are merely
allowed (may). In an MTS every required transition is also allowed, to avoid
inconsistencies.  An MTS describes simultaneously an over-approximation and an
under-approximation of a process in an intertwined manner.  In a stepwise
refinement scenario this approximation interval is narrowed down to a single
process, an LTS.

Subsequent work has lifted the assumption that required transitions need
also be allowed, leading to the model of mixed transition systems
\cite{dams96}.  
This means that mixed 
transition systems allow states to have requirements that are impossible to
fulfill, which we will refer to as \emph{conflicting} requirements. 

However, the general notion of a must transition that is not also a may transition
appears quite intricate; it calls for interpreting for each action the specifications at the
targets of the must transitions with that action, which must \emph{all} be satisfied in
conjunction with some choice of may transition with that action. This is  
reflected in the complexity of basic implementability being EXPTIME complete
for Mixed transition systems, and trivial for MTSs\,\cite{DBLP:journals/mscs/AntonikHLNW10}.

MTS design focuses on underspecification of behaviour that is resolved at \emph{implementation time}. So for each implementation of a MTS, allowed behaviour has to be refined to some desirable subset, in agreement with the must behaviour.  While this is the main carrying feature of MTSs, it does  
side-step 
the dual problem of under-specifying behaviour at \emph{runtime}, such that allowed (may) behaviours are disambiguated during execution of a process, and required (must) behaviours are performed at runtime to ensure progress (liveness). 

In the present paper, we propose taking a step back to consider a model for specification of may and must behavior both at run-time (as acceptance crieteria on execution traces) and at implementation time (as refinements).
When considering an MTS as a model for underspecified behavior to be resolved at 
run-time leads to a natural notion of accepting and deadlocked states. An accepting state 
is a state from which no must transitions are specified, and thus it is permitted to end the 
execution. A deadlocked state is a state from which no may transitions are specified but at least one must transition is specified.

Motivated by the response pattern for property specifications~\cite{Dwyer99patternsin} and previous work on declarative workflow modelling languages~\cite{places2010,sefm:2011}, 
we propose to interpret
 must transitions as a requirement that an action within a certain scope must \emph{eventually} be executed. The \emph{scope} is the extend of the execution in which a must transition with that activity is defined.
 This leads to a new, relaxed interpretation of states of mixed transition systems having must transitions that are not allowed in the same state as a may transition: Such states are only expressing a conflict, i.e. unacceptable, behavior at run-time if the scope of the must transition includes the rest of the execution, and the activity required by the must transition is not executed. In other words, a run will be considered unacceptable if an activity is continuously required as a must transition but never executed. For finite trace semantics, which we will focus on in the present paper, this condition is implied by the final state being an accepting state as defined above. For infinite trace semantics, which we will study in future work as indicated in Sec.~\ref{sec:future}, this condition corresponds to the \emph{future} modality of linear-time temporal logic, and allows $\omega$-regular languages to be expressed in a natural way.

While it thus makes sense to consider must behavior that is not also (immediately) allowed as may behavior, we do not want to unleash the full generality of mixed transition systems. Instead we propose to
replace the must transitions by a set of
\emph{response actions} assigned to 
each state,
referred to as the \emph{response set}.
The intuitive meaning is that if an action belongs to the response set of a state $s$, then the action is
required 
before termination, unless a state is reached in which the action no longer belongs to the response set. 
We name the
resulting model \emph{Transition Systems with Responses} (TSR).  

We show that TSRs correspond to a restricted class of mixed transition systems that we refer to as action-deterministic mixed transition systems, and transfer the standard definition of refinement for mixed (and modal) transition systems.
However, a simple example shows that the standard definition of refinement does not preserve deadlock freedom. Consequently, we instead
propose studying \emph{safe} refinements, which are those refinements that reflect deadlocked states (i.e., preserve deadlock freedom).

\newcommand{\figscale}{0.7}

\begin{figure}[t]
\centering
\subfigure[Medication workflow]{\label{fig:medication}\includegraphics[width=70mm]{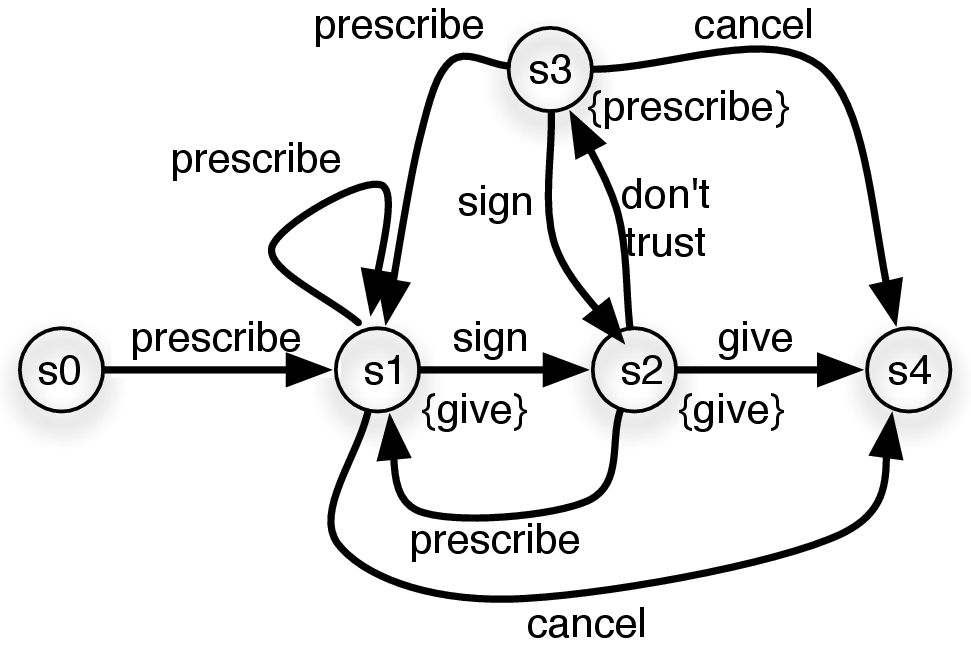}}    
\subfigure[Safe refinement]{\label{fig:refinement}\includegraphics[scale=\figscale]{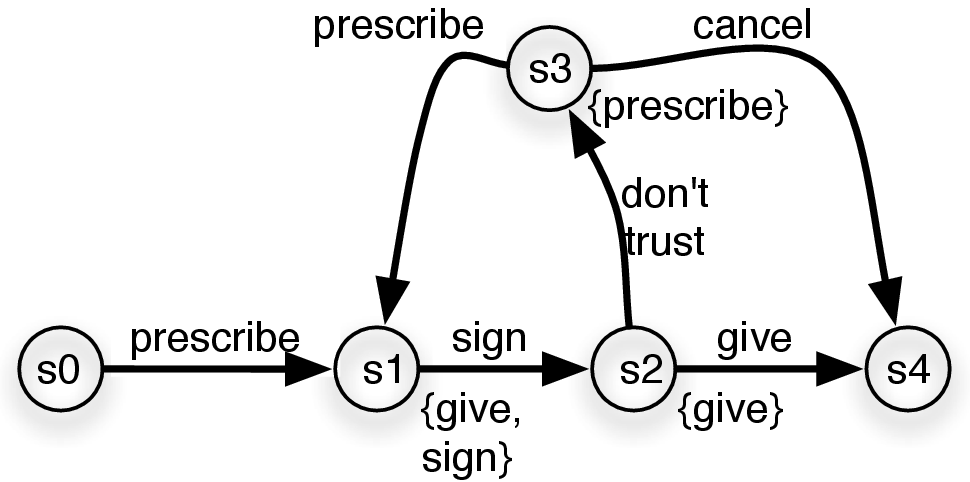}}\subfigure[Unsafe refinement]{\label{fig:unsaferefinement}
\includegraphics[scale=\figscale]{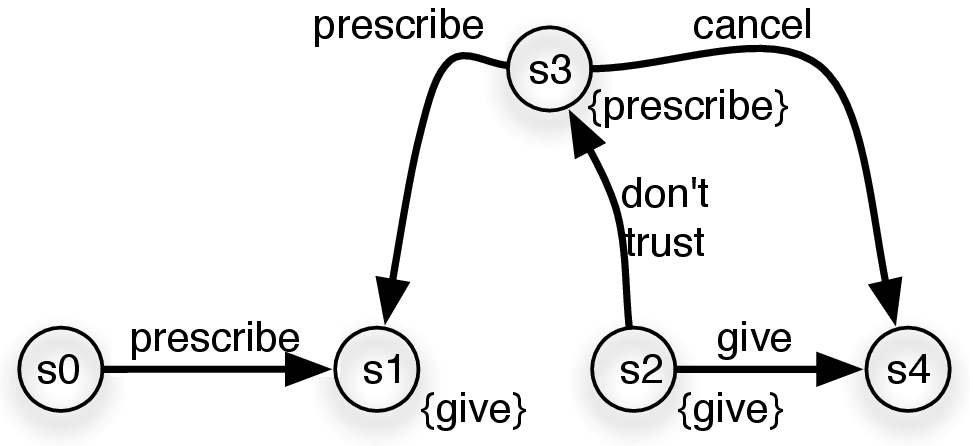}}
\caption{Medication workflow as TSR and refinements.\label{fig:medicationexamples}}
\end{figure}

As an example, consider the TSR given in Fig.~\ref{fig:medication} with initial state \textsf{s0}. It describes a medication workflow in which a doctor prescribes medicine, possible several times, and then either cancels or signs the prescription. A prescription has as the required response that a nurse gives the medicine. This will end the workflow and cannot happen if the doctor has not signed after the latest prescription. However, the nurse may instead indicate that the prescription is not trusted. In that case, the doctor is requested as response to prescribe new medicine, but may instead just sign the old prescription (indicating that it was indeed right) or cancel the prescription. 

The TSR in Fig.~\ref{fig:refinement} shows a refinement of the TSR in Fig.~\ref{fig:medication}. The refinement restricts the workflow such that only one prescription can be made at a time, by removing some of the prescription transitions that are not required. Also, the doctor is now \emph{required} to sign after making a prescription, i.e., as a response. This is enforced by adding \textsf{sign} to the response set of state \textsf{s1}, corresponding to making the \textsf{sign} transition from state \textsf{s1} a must transition.
Moreover, the doctor can now only cancel if a prescription is indicated as not trusted. This refinement is \emph{safe}, since it does not introduce any deadlocked states. Moreover, as will be more clear below when we give the formal definitions, the TSR in  Fig.~\ref{fig:refinement} has no unsafe refinements.
The TSR in Fig.~\ref{fig:unsaferefinement} un the other hand shows an unsafe refinement of the TSR in  Fig.~\ref{fig:medication} since it introduces a deadlocked state. The state \textsf{s1} is deadlocked since \textsf{give} is required but no transition is possible.
\begin{figure}[t]
\centering
\includegraphics[scale=\figscale]{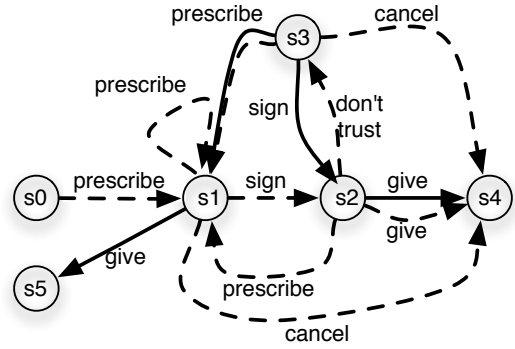}
\caption{Medication workflow as Mixed TS.\label{fig:mixedmedication}}
\end{figure}
Finally,
Fig.~\ref{fig:mixedmedication} shows a mixed transition system corresponding to the TSR in Fig.~\ref{fig:medication}. As usual, the solid transitions are must transitions and dashed transitions are may transitions. Note that for all must transitions we have a corresponding may transition, except for the must transition from
state \textsf{s1} to state \textsf{s5}. This transition captures that it in state \textsf{s1} is required (eventually) to give the medicine, unless another transition causes the must transition to dissapear, as for instance the \textsf{don't trust} transition from \textsf{s2} to \textsf{s3}.

The rest of the paper is structured as follows. In Sec.~\ref{sec:modalmixed} we briefly recall the definition of mixed and modal transition systems and refinement for such, and define the restricted classes of action-deterministic modal and mixed transition systems. In Sec.~\ref{sec:tsr} we then give the formal definition of transition systems with responses, and safe and unsafe refinement for such, prove the correspondence to action-deterministic modal and mixed transition systems and relate refinement to language inclusion. In Sec.~\ref{sec:future} we conclude and provide pointers to future work.

%% file: modalmixed.tex
In this section we briefly recall the definition of  modal and mixed transition systems, define the subclasses of action-deterministic modal and mixed transition systems and  recall the standard definition of refinement for modal and mixed transition systems.

\begin{definition}[Mixed and Modal Transition Systems]
A \emph{Mixed Transition System} (MixTS) is a tuple $T = \langle S, s_0,
\actionSet,  ->_{\must},   ->_{\may} \rangle$ where $S$ is a set of
\emph{states}, $s_0 \in S$ is the \emph{initial state}, $\actionSet$ is a set
of actions, and $->_{\must}, ->_{\may} \subseteq S \times \actionSet \times S$
are respectively \emph{must} and \emph{may} transition relations.
$T$ is also a \emph{Modal Transition System} (MTS) if additionally $->_{\must}
\,\, \subseteq \,\, ->_{\may}$. Finally, let $s\action{a}s'$ denote that either $s\action{a}_{\may}s'$ or $s\action{a}_{\must}s'$.
\end{definition}

We will in this paper focus on so-called \emph{action-deterministic} mixed and modal transition systems that we will show directly correspond to the proposed model of transition systems with responses. 
\begin{definition}[Action-deterministic MixTS and MTS]
A (mixed, modal) transition system $T = \langle S, s_0,
\actionSet,  ->_{\must},   ->_{\may} \rangle$ is \emph{action-deterministic} if 
\begin{enumerate}

\item $s\action{a}s' \wedge s\action{a}s''$ implies $s'=s''$.
\item $s\action{a}_{\must}s' \wedge s\not\action{a}_{\may}s'$ 
implies $\forall a\in\actionSet. s'\not\action{a}$ and $s''\action{b}s'$ implies $s=s''\wedge a=b$. 
\end{enumerate}
\end{definition}

The first condition is the usual determinacy condition, stating that for any  state, the target state for a transition with a specific action is unique.
However, note that since we do not distinguish between may and must transitions, it also implies that if there is a may and must transition with the same label from a transition they lead to the same state.
The second condition restricts the occurrence of must transitions that have no corresponding may transition, and is thus trivially satisfied for modal transition systems. The condition states that such a "must but may not" transition must  lead to a state from which no further transitions are possible, and that it is the unique transition leading to that state.
Intuitively, the restriction means that such a must transition contains no other information than the fact that the action is required. This allows us in the next section to replace must transitions with sets of actions assigned to each state in the definition of transition systems with responses.

Below we recall the definition of refinement for mixed and modal TS.

\begin{definition}[Refinement for MixTS and MTS]
  A binary relation $\mathcal R \subseteq S_1\times S_2$ between the state sets of two mixed transition systems   $T_j = <. S_j,i_j,
  \actionSet, ->_{\must j},   ->_{\may j} .>$ for $j\in\{1,2\}$    is 
  a
  refinement if 
  \begin{enumerate}
  \item $i_1 \mathcal R i_2$ and 
   \item $s_1 \mathcal R s_2$ implies
  \begin{enumerate}
      \item $\forall s_1 \action{a}_{\must 1} s'_1$ 
            implies $\exists s_2 
       \action{a}_{\must 2} s'_2$, 
        and 
        $s'_1 \mathcal R s'_2$,
                \item 
          $\forall s_2 \action{a}_{\may 2} s'_2$
          implies $\exists s_1 \action{a}_{\may 1} s'_1$ and $s'_1 \mathcal R
          s'_2$
  \end{enumerate}
 \end{enumerate}

\end{definition}

Since identities are refinements and refinements compose as relations to refinements we get categories \cat{MTS} and \cat{MixTS}, having respectively modal and mixed transition systems as objects and refinements as arrows, and the two sub categories \cat{DMTS} and \cat{DMixTS} induced by action-deterministic modal and mixed transition systems respectively.

%% file: tsr.tex
As described in the introduction, the definition of transition system with responses replaces the must transition relation with a set of response actions for each state. 

\begin{definition}%
A \emph{Transition System with Responses} (TSR) is a tuple $T = \langle S,
s_0, \actionSet,  \must,    -> \rangle$ where $S$, $s_0$, $\actionSet$  are
like above and 
   $-> \subseteq S \times \actionSet \times S$ is an action-deterministic transition
          relation and
          $\must : S -> \power(\actionSet)$ 
          defines for each state the \emph{response actions}.
 Let $\may(s)=_{def}\{a\mid \exists s'. s\action{a}s'\}$, i.e., the actions on transitions that may be taken from $s$.   
We then say that a TSR is \emph{modal} if $\forall s\in S. \must(s)\subseteq \may(s)$. 
          \end{definition}

As stated by the proposition below, the class of TSRs are in bijective correspondence (up to graph isomorphism) with action-deterministic mixed transition systems. As usual we let \obj{C} refer to the class of objects of a category \cat{C}.

\begin{proposition}[Representation of TSRs as DMixTSs]
\label{prop:representation}
There are maps $MR: \DMixTS-> \TSR$ and $RM:\TSR-> \DMixTS$ such that for any DMixTS $M$, $RM(MR(M))$ is isomorphic to $M$ and for any TSR $T$, $MR(RM(T))=T$.
\begin{proof}(outline)
An action-deterministic mixed transition system $M = \langle S, s_0, \actionSet,
->_{\must},    ->_{\may} \rangle$ has as corresponding TSR $MR(M) =
\langle S, s_0, \actionSet,\linebreak[0]  \must,    -> \rangle$ where $\must(s)=_{def}\{a\mid \exists s'\in S . s\action{a}_{\must}s'\}$ and $-> = ->_{\may}$. Conversely, a TSR $T= \langle S, s_0, \actionSet,  \must,    -> \rangle$ has as corresponding action-deterministic MixTS $RM(T)=\langle S\cup\{s_{\must a}\mid s\in S \wedge \must(s)\backslash\may(s)\not=\emptyset\}, s_0, \actionSet,  ->_{\must},    ->_{\may} \rangle$,  
where $->_{\must} = \{(s,a,s')\mid a\in\must(s) \wedge (s\action{a}s' \vee (s\not\action{a} \wedge s'=s_{\must a})\}$ and $->_{\may} = ->$, where we assume $\forall a\in\actionSet. \forall s\in S. s_{\must a}\not\in S$.
\end{proof}
\end{proposition}

The key idea of the representation of action-deterministic mixed transition systems as TSRs, exemplified by Fig.~\ref{fig:medication} and Fig.~\ref{fig:mixedmedication}, is that we forget about the destination state of must transitions and simply record the presence of a must transition with action $a$ from a state \textsf{s} by an action $a$ in the response set of \textsf{s}. The action determinacy conditions ensure that we can recover the original mixed transition system (up to graph isomorphism). For the map in the other direction, note that in the case where an action in the response set has no corresponding may transition we introduce a new state $\mathsf{s}_{\must a}$ as destination of the must transition. This explains the state $s5$ in Fig.~\ref{fig:mixedmedication} which arises as the state $\mathsf{s1}_{\must \mathsf{give}}$.

It is easy to see that the correspondence in the proposition above restricts to a correspondence between modal TSR and action-deterministic modal TS.
\begin{proposition}[Representation of Modal TSRs as DMTSs]
\label{prop:modalrepresentation}
The maps $MR: \DMixTS-> \TSR$ and $RM:\TSR-> \DMixTS$ restricts to maps $MR: \DMTS-> \MTSR$ and $RM:\MTSR-> \DMTS$.
\end{proposition}

As mentioned in the introduction, we interpret actions in the response sets as actions that during an execution must \emph{eventually} be executed or be excluded from the response set. If we only consider finite trace semantics, this interpretation makes it natural to define accepting states for TSR as states with an empty response set, allowing us to use the standard definition of the language for a finite automaton to define the language for a TSR.

\begin{definition}[Language of a TSR]
      We refer to a finite 
          sequence of transitions $s_0\action{a_0}s_1\action{a_1}\ldots\action{a_{n-1}}s_{n}$ starting at
	  the initial state as a \emph{run} and define that it is \emph{accepting}, if $\must(s_n)=\emptyset$. 
	  We let $L(T)$, referred to as the \emph{language} of $T$, denote the set of all action sequences labelling accepting runs.
\end{definition}
As an example, the regular language of the medication workflow in Fig.~\ref{fig:refinement} is given by the expression
expression $\epsilon + \textsf{prescribe}. \textsf{sign} (\textsf{don't trust}. \textsf{prescribe}. \textsf{sign})^*(\textsf{don't trust}.\textsf{cancel} + \textsf{give})$.

In addition to accepting states, we may naturally define deadlocked states as states with a non-empty response set but no out-going transitions. 

\begin{definition}[Deadlocked state]
A \emph{deadlocked} state in a TSR $T = \langle S, s_0, \actionSet,  \must,    -> \rangle$ is a state with a non-empty \emph{must} set, and no out-going transitions, i.e., a state in which some actions are required but no further transitions are possible. Formally 
we define a predicate $\deadlock$ on $S$ by
$\deadlock(s) \equiv \must(s) \neq \emptyset \land \may(s) = \emptyset$. A TSR is \emph{deadlock free} if it has no reachable deadlock state.
\end{definition}

As already mentioned in the introduction, the state \textsf{s1} is deadlocked in the TSR in Fig.~\ref{fig:unsaferefinement} since it has the non-empty response set $\{\mathsf{give}\}$ but no out-going transitions.
It follows trivially from the definition that a modal TSR can not have any deadlock states.

\begin{lemma}
Every modal TSR is deadlock free.
\begin{proof}
For any state $s$ of a modal TSR it holds by definition that $\must(s)\subseteq \may(s)$, so it can not be the case that $\must(s) \neq \emptyset \land \may(s) = \emptyset$
\end{proof}
\end{lemma}

We now define refinement for TSR in Def.~\ref{def:refinementTSR} below, guided by the representation of TSRs as DMixTS given in Prop.~\ref{prop:representation}. Condition~\ref{cond:refine_response} ensures that states in the refined system require at least the same responses as the states they refine. Condition~\ref{cond:refine_preserve} ensures that transitions with actions required as responses are preserved by the refinement, which by condition~\ref{cond:refine_response} also will be required as responses.  Finally, condition~\ref{cond:refine_reflect} ensures that the refined system do not introduce transitions that can not be matched in the system being refined.
We then define \emph{safe} refinements as refinements satisfying the extra condition that deadlock states must be reflected.
\begin{definition}[Refinement]
\label{def:refinementTSR}
  A binary relation $\mathcal R \subseteq S_1\times S_2$ between the state sets of two transition systems with responses   $T_j = <. S_j,i_j,
  \actionSet, \must_j,   ->_j .>$ for $j\in\{1,2\}$    is 
  a
  refinement if 
  \begin{enumerate}
  \item $i_1 \mathcal R i_2$ and 
  \item  $s_1 \mathcal R s_2$ implies
   \begin{enumerate}
      \item $\must(s_1)\subseteq \must(s_2)$, \label{cond:refine_response} 
      \item $\forall s_1 \action{a}_1 s'_1$ and $a\in\must_1(s_1)$ implies $\exists s_2 
       \action{a}_2 s'_2$ 
        and 
        $s'_1 \mathcal R s'_2$,\label{cond:refine_preserve} 
        \item           $\forall s_2 \action{a}_2 s'_2$
          implies $\exists s_1 \action{a}_1 s'_1$ and $s'_1 \mathcal R
          s'_2$\label{cond:refine_reflect} 
    \end{enumerate}
 \end{enumerate}
The refinement $\mathcal R$ is safe if it satisfies the additional condition that $s_1\mathcal
R s_2$ implies $\deadlock(s_2) \implies \deadlock(s_1)$, i.e. 
it reflects deadlock states.
\end{definition}
Looking at the example TSRs in Fig.~\ref{fig:medicationexamples}, which we will refer to as $T_a$, $T_b$, and $T_c$, it is easy to verify that the identity relation on states is a refinement from $T_a$ to $T_b$ and from $T_a$ to $T_c$.  Also, it is easy to see that here is no refinement between $T_b$ and $T_c$: Any refinement must relate the initial states and thus also states \textsf{s1} to each other because of condition~\ref{cond:refine_reflect}. Now, $T_b$ cannot refine $T_c$ since it has a transition from \textsf{s1} which is not matched in $T_c$ as required by condition~\ref{cond:refine_reflect}. Conversely, $T_c$ cannot refine $T_b$ since it does not match the \textsf{sign} transition from \textsf{s1}, which is required by condition~\ref{cond:refine_preserve} since it is in the response set of \textsf{s1} in $T_b$. Also, the response set of \textsf{s1} in $T_b$ is not included in the response set of \textsf{s1} in $T_c$ as required by 
condition~\ref{cond:refine_response}. Finally, the refinement from $T_a$ to $T_c$ is not safe since state \textsf{s1} is deadlocked in $T_c$ as mentioned above, and this is not the case in $T_a$ since the \textsf{sign} transition is possible.

Since identities are refinements and refinements compose as relations to refinements we get a category $\mathbf{TSR}$ with TSRs as objects and refinements as arrows.
As shown in the following proposition, a refinement between two DMixTSs is also a refinement of the corresponding TSRs and vice versa, thus the maps given in Prop.~\ref{prop:representation} extends to a functor.

\begin{proposition}
The map $MR: \DMixTS-> \TSR$ extends to a functor from $\cat{DMixTS}$ to $\cat{TSR}$.
\begin{proof}
It follows from the definition that for two DMixTS $M_i=\langle S_i,s_{i,0},\actionSet,\must_i,->_i \rangle$ for $i\in\{1,2\}$,  if $ \mathop{\mathcal{R}}\subseteq S_1 \times S_2$ is a refinement then it is also a refinement between the corresponding TSRs $MR(M_i)$.
\end{proof}
\end{proposition}

Conversely, a refinement between two TSRs can be mapped to the corresponding DMixTS, extending the map $RM$ to a functor.

\begin{proposition}
The map $RM: \TSR -> \DMixTS$ extends to a functor from  $\cat{TSR}$ to $\cat{DMixTS}$.
\begin{proof}

Given two TSRs $T_i= \langle S_i, s_{i,0}, \actionSet,  \must_i,    ->_i \rangle$ for $i\in\{1,2\}$, if $ \mathop{\mathcal{R}}\subseteq S_1 \times S_2$ is a refinement then $\mathop{\mathcal{R'}}\subseteq S'_1 \times S'_2$ is a refinement for the corresponding DMixTS $RM(T_i)=\langle S'_i,s_{i,0},\actionSet,\must_i,->_i \rangle$, where $\mathop{\mathcal{R'}}
= \mathop{\mathcal{R}}\cup \{(s_{\must a},s'_{\must a})\mid s \mathop{\mathcal{R}} s' \}$.
\end{proof}
\end{proposition}

It follows that the definition of refinement for TSRs correspond to refinement between the corresponding DMixTS. 
\begin{theorem}
The category $\mathbf{TSR}$  is equivalent to the category $\mathbf{DMixTS}$.
\end{theorem}

The theorem above shows that TSRs and refinement are indeed equivalent to action-deterministic mixed transition systems with refinement.
However, it is easy to give an example of a refinement which is not safe, as illustrated by the TSR in Fig.~\ref{fig:unsaferefinement}.
This suggests that we should really work in the sub category of TSR with safe refinements.

Below we prove that refinement for TSRs implies language inclusion.
\begin{proposition}
Given two TSRs $T_i= \langle S_i, s_{i,0}, \actionSet,  \must_i,    ->_i \rangle$ for $i\in\{1,2\}$.
If there exists a refinement $ \mathop{\mathcal{R}}\subseteq S_1 \times S_2$ 
then $L(T_2)\subseteq L(T_1)$. \begin{proof}
Assume $ \mathop{\mathcal{R}}\subseteq S_1 \times S_2$ is a refinement of two TSRs $T_i= \langle S_i, s_{i,0}, \actionSet,  \must_i,    ->_i \rangle$ for $i\in\{1,2\}$ and $\sigma_2=s_{2,0}\action{a_0}s_{2,1}\action{a_1}\ldots\action{a_{n-1}}s_{2,n}$ is an accepting run of $T_2$. Then by condition~\ref{cond:refine_reflect} of Def.~\ref{def:refinementTSR} there exists a run $\sigma=s_{1,0}\action{a_0}s_{1,1}\action{a_1}\ldots\action{a_{n-1}}s_{1,n}$ of $T_2$ such that $s_{1,i} \mathop{\mathcal{R}} s_{2,i}$. By condition~\ref{cond:refine_response} and $\must(s_{2,n})=\emptyset$ it then follows that   $\must(s_{1,n})=\emptyset$, so the run $\sigma_1$ is accepting and has the same sequence of actions as $\sigma_2$.
\end{proof}
\end{proposition}
\begin{figure}[t]
\centering
\includegraphics[scale=\figscale]{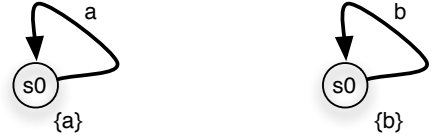}
\caption{Example of TSRs with empty languages but not refining each other.\label{fig:counterexample}}
\end{figure}

The converse, that language inclusion implies refinement, is not true.
As a counter example, consider the two TSRs in Fig.~\ref{fig:counterexample} with a single transition each from the initial state to the initial state, but with two different actions, and such that the action is a member of the response set. Both TSRs have empty language (and no deadlocks) but is not related by refinement in any direction. One may however argue, that this counter-example is an artefact of restricting attention to finite trace semantics. Indeed, the infinite sequence of \textsf{a} actions should be acceptable in the left system and the infinite sequence of \textsf{b} actions should be acceptable in the right system since it fulfils the response constraint. We leave for future work to investigate TSRs and refinement for infinite trace semantics, and if language inclusion in this case does indeed imply refinement for TSRs.

%% file: conclusion.tex
We have introduced Transition Systems with Responses (TSRs) as a conservative generalisation of action-deterministic Modal Transition Systems, which allows for static (implementation time) refinements as well as dynamic (runtime) resolution of underspecified behaviour. We have proven that the TSR model corresponds to a restricted class of mixed transition systems, which we refer to as the \emph{action-deterministic} mixed transition systems. This class of mixed transition systems is much simpler than general mixed transition systems, and yet allows for a natural definition of deadlocks. 
We have formulated the standard refinement for mixed TS in terms of TSRs and proposed studying \emph{safe} refinements, which are refinements that reflect deadlocked states, i.e., those which preserve deadlock freedom.

We leave as future work the study of how to lift the restriction to action-deterministic MTS, and how to treat infinite computations and liveness for TSR and refinements for such.